\documentclass[12pt]{article}

\usepackage[centertags]{amsmath}
\usepackage{amsfonts}
\usepackage{amssymb}
\usepackage{amsthm}
\usepackage{newlfont}
\usepackage{epsfig}
\usepackage{amscd}

\newcommand{\RR}{{\mathbb R}}
\newcommand{\ZZ}{{\mathbb Z}}

\newcommand{\CC}{{\mathbb C}}
\newcommand{\beq}{\begin{equation}}
\newcommand{\eeq}{\end{equation}}
\newcommand{\ba}{\begin{array}}
\newcommand{\ea}{\end{array}}
\newcommand{\bea}{\begin{eqnarray}}
\newcommand{\eea}{\end{eqnarray}}

\begin{document}

\begin{center}
{\large \sc \bf The dispersionless 2D Toda equation: dressing,} 

\vskip 10pt 
{\large \sc \bf Cauchy problem, longtime behaviour,} 

\vskip 10pt 

{\large \sc \bf  implicit solutions and wave breaking}

\vskip 20pt

{\large  S. V. Manakov$^{1,\S}$ and P. M. Santini$^{2,\S}$}

\vskip 20pt

{\it 
$^1$ Landau Institute for Theoretical Physics, Moscow, Russia

\smallskip

$^2$ Dipartimento di Fisica, Universit\`a di Roma "La Sapienza", and \\
Istituto Nazionale di Fisica Nucleare, Sezione di Roma 1 \\
Piazz.le Aldo Moro 2, I-00185 Roma, Italy}

\bigskip

$^{\S}$e-mail:  {\tt manakov@itp.ac.ru, paolo.santini@roma1.infn.it}

\bigskip

{\today}

\end{center}

\begin{abstract}
We have recently solved the inverse spectral problem for one-parameter families   
of vector fields, and used this result to construct the formal solution of the Cauchy 
problem for a class of integrable nonlinear partial differential equations in multidimensions,  
including the second heavenly equation of Plebanski and the dispersionless Kadomtsev 
- Petviashvili (dKP) equation, arising as commutation of vector fields. 
In this paper we make use of the above theory i) to construct the nonlinear Riemann-Hilbert dressing 
for the so-called two dimensional dispersionless Toda equation 
$\left(exp(\varphi)\right)_{tt}=\varphi_{\zeta_1\zeta_2}$,  
elucidating the spectral mechanism responsible for wave breaking; ii) we present the formal solution of the Cauchy 
problem for the wave form of it: $\left(exp(\varphi)\right)_{tt}=\varphi_{xx}+\varphi_{yy}$;  
iii) we obtain the longtime behaviour of the solutions of such a Cauchy problem, showing 
that it is essentially described by the longtime breaking formulae of the dKP solutions,   
confirming the expected universal character of the dKP equation as prototype model in the description 
of the gradient catastrophe of two-dimensional waves; iv) we finally characterize a 
class of spectral data allowing one to linearize the RH problem, corresponding to a class of implicit 
solutions of the PDE.    
 
\end{abstract}
%%%%%%%%%%%%%%%%%%%%%%%%%%%%%%%%%%%%%%%%%
\section{Introduction}

It was observed long ago \cite{ZS} that the commutation of multidimensional 
vector fields can generate integrable nonlinear partial differential equations (PDEs) in arbitrary 
dimensions. Some of these equations are dispersionless (or quasi-classical) limits 
of integrable PDEs, having the dispersionless Kadomtsev 
- Petviashvili (dKP) equation $u_{xt}+u_{yy}+(uu_x)_x=0$ \cite{Timman},\cite{ZK} as universal prototype example, 
they arise in various problems of 
Mathematical Physics and are intensively studied in the recent literature 
(see, f.i., \cite{KG} - \cite{KM}). In particular, an elegant integration scheme applicable, in general,  
to nonlinear PDEs associated with Hamiltonian vector fields, was presented in \cite{Kri} and a 
nonlinear $\bar\partial$ - 
dressing was developed in \cite{K-MA-R}. Special classes of nontrivial solutions were also derived 
(see, f.i., \cite{DT}, \cite{G-M-MA}).

The Inverse Spectral  
Transform (IST) for $1$-parameter families of multidimensional vector fields has been 
developed in \cite{MS1} (see also \cite{MS2}). This theory, introducing interesting novelties 
with respect to the classical IST for soliton equations \cite{ZMNP,AC}, 
has allowed one to construct the formal solution of the Cauchy problems for the second   
heavenly equation  \cite{Pleb} in \cite{MS1} and for the novel system of PDEs
\beq
\label{dKP-system}
\ba{l}
u_{xt}+u_{yy}+(uu_x)_x+v_xu_{xy}-v_yu_{xx}=0,         \\
v_{xt}+v_{yy}+uv_{xx}+v_xv_{xy}-v_yv_{xx}=0,
\ea
\eeq
in \cite{MS3}. The Cauchy problem for the $v=0$ reduction of (\ref{dKP-system}), the dKP equation, was also presented 
in \cite{MS3}, while the Cauchy problem for the 
$u=0$ reduction of (\ref{dKP-system}), an integrable system introduced 
in \cite{Pavlov}, was given in \cite{MS4}. This IST and its associated nonlinear Riemann - Hilbert (RH) dressing 
turn out to be efficient 
tools to study several properties of the solution space, such as: i) the characterization of a 
distinguished class of spectral data for which the associated nonlinear RH problem is linearized and solved, 
corresponding to a class of implicit solutions of the PDE (as it was done for the dKP equation in \cite{MS5}
 and for the Dunajski 
generalization \cite{Duna} of the second heavenly equation in \cite{BDM}); ii) the construction of the 
longtime behaviour of the solutions of the Cauchy problem \cite{MS5};  
iii) the possibility  to establish  whether or not the lack of dispersive terms in the nonlinear PDE 
causes the breaking of localized initial profiles and, if yes, to investigate in a surprisingly explicit way the analytic 
aspects of such a wave breaking (as it was recently done for the (2+1)-dimensional dKP model in \cite{MS5}). 
 
In this paper we make use of this theory to study another distinguished model arising as the commutation of 
vector fields, the so-called 2 dimensional dispersionless Toda (2ddT) equation
\beq\label{dT1}
\phi_{\zeta_1\zeta_2}=\left(e^{\phi_t}\right)_t,~~~\phi=\phi(\zeta_1,\zeta_2,t),   
\eeq
or 
\beq\label{dT2}
\varphi_{\zeta_1\zeta_2}=\left(e^{\varphi}\right)_{tt},~~~~\varphi=\phi_t,
\eeq
also called Boyer-Finley \cite{BF} equation or SU($\infty$) Toda equation \cite{Ward}, the natural continuous 
limit of the 2 dimensional Toda lattice \cite{Darboux,Mikhailov} 
\beq
{\phi_n}_{\zeta_1\zeta_2}=c\left(e^{\phi_{n+1}-\phi_{n}}-e^{\phi_{n}-\phi_{n-1}}\right),~~\phi=\phi_n(\zeta_1,\zeta_2).
\eeq
  
The 2ddT equation was probably first derived in \cite{FP} as an exact reduction of the second heavenly equation; then in 
\cite{Zak} as a distinguished example of an integrable system in multidimensions.   
Some of its integrability properties have been investigated in \cite{Saveliev} and the integration method presented in 
\cite{Kri} is applicable to it.  
Both elliptic and hyperbolic versions of (\ref{dT1}) are relevant, describing, for instance,  
integrable ${\cal H}$-spaces (heavens) 
\cite{BF,GD} and integrable Einstein - Weyl geometries \cite{H}-\cite{J},\cite{Ward}.  
String equations solutions \cite{TT2} of it are relevant in the ideal Hele-Shaw problem 
\cite{MWZ,WZ,KMZ,LBW,MAM}.   

The integrability of (\ref{dT1}) follows from the fact that (\ref{dT1}) is the condition of commutation 
$[\hat L_1,\hat L_2]=0$ for the following pair of one-parameter families of vector fields \cite{TT1}:
\beq\label{VF1}
\ba{l}
\hat L_1=\partial_{\zeta_1}+\lambda v\partial_{t}+
\left(-\lambda v_t+\frac{\phi_{\zeta_1 t}}{2}\right)\lambda\partial_{\lambda} ,\\ 
\hat L_2=\partial_{\zeta_2}+\lambda^{-1} v\partial_{t}+
\left(\lambda^{-1}v_t-\frac{\phi_{\zeta_2 t}}{2}\right)\lambda\partial_{\lambda}, 
\ea
\eeq 
where
\beq
\ba{l}
%z=\frac{x+iy}{2},~~\bar z=\frac{x-iy}{2}
v=e^{\frac{\phi_{t}}{2}}
\ea
\eeq 
and $\lambda\in\CC$ is the spectral parameter, implying the existence of common eigenfunctions of both vector fields; 
i.e., the existence of the Lax pair: 
\beq\label{Lax1}
\ba{l}
\psi_{\zeta_1}=-\lambda v \psi_t+
\left(\lambda v_t-\frac{\phi_{\zeta_1 t}}{2}\right)\lambda\psi_{\lambda}, \\
\psi_{\zeta_2}=-\lambda^{-1} v \psi_t -
\left(\lambda^{-1}v_t-\frac{\phi_{\zeta_2 t}}{2}\right)\lambda\psi_{\lambda} .
\ea
\eeq

Equations (\ref{Lax1}) and (\ref{dT1}) can be written in the following ''Hamiltonian'' form \cite{TT1}:  
\beq\label{Ham1_Lax}
\ba{l}
\psi_{\zeta_1}+\{{\cal H}_1,\psi \}_{(\lambda ,t)}=0, ~~
\psi_{\zeta_2}+\{{\cal H}_2,\psi \}_{(\lambda ,t)}=0, \\
{{\cal H}_1}_{\zeta_2}-{{\cal H}_2}_{\zeta_1}-\{{\cal H}_1,{\cal H}_2 \}_{(\lambda ,t)}=0,
\ea
\eeq 
where 
\beq\label{Ham1}
\ba{l}
{\cal H}_1= \lambda v-\frac{\phi_{\zeta_1}}{2},~~{\cal H}_2=-\lambda^{-1}v+\frac{\phi_{\zeta_2}}{2}
\ea
\eeq 
and 
\beq\label{PB1}
\{f,g\}_{(\lambda ,t)}:=\lambda (f_{\lambda} g_t-f_t g_{\lambda}).
\eeq

If, in particular, 
\beq\label{elliptic}
\zeta_1=z=\frac{x+iy}{2},~~\zeta_2=\bar z=\frac{x-iy}{2},~~~x,y\in\RR,
\eeq
equation (\ref{dT1}) becomes the following nonlinear wave equation
\beq\label{dT3}
\phi_{xx}+\phi_{yy}=\left(e^{\phi_t}\right)_{t},
\eeq
or
\beq\label{dT4}
\varphi_{xx}+\varphi_{yy}=\left(e^{\varphi}\right)_{tt}.
\eeq
In addition, if $|\lambda|=1$ and $\phi\in\RR$, equations (\ref{Ham1}) give the ``real'' Hamiltonian 
formulation \cite{Zak} 
\beq\label{Ham2_Lax}
\ba{l}
\psi_x+\{H_1,\psi \}_{(\theta ,t)}=0, ~~
\psi_{y}+\{H_2,\psi \}_{(\theta ,t)}=0, \\
{H_1}_y-{H_2}_x-\{H_1,H_2 \}_{(\theta ,t)}=0,
\ea
\eeq 
of (\ref{Lax1}) and (\ref{dT3}), for the real Hamiltonians $H_1,H_2$:
\beq\label{Ham_real}
\ba{l}
H_1=\frac{i}{2}({\cal H}_1+{\cal H}_2)=\sin\theta e^{\frac{\phi_{t}}{2}}-\frac{1}{2}\phi_y,  \\
H_2=\frac{1}{2}({\cal H}_2-{\cal H}_1)=-\cos\theta e^{\frac{\phi_{t}}{2}}+\frac{1}{2}\phi_x
\ea
\eeq 
and the Poisson bracket
\beq\label{PB_real}
\ba{l}
\{f,g \}_{(\theta ,t)}:=f_{\theta}g_t-f_tg_{\theta},
\ea
\eeq 
having introduced the parametrization
\beq\label{param_lambda}
\lambda = e^{-i\theta},~~\theta\in\RR .
\eeq

The paper is organized as follows. In \S 2 we present the dressing scheme for equations (\ref{dT1}) and 
(\ref{dT3}), given in terms of a vector nonlinear RH problem. As for the dressing of dKP presented in \cite{MS5}, 
since the normalization of the eigenfunctions turns out to depend  
on the unknown solution of 2ddT, a closure condition is necessary, allowing one to construct the solution of 2ddT 
through an implicit system of algebraic equations, whose inversion is responsible for the wave breaking of 
an initial localized profile. In \S 3 we present the IST for the 2ddT equation 
(\ref{dT3}) and use it to obtain the formal solution of the Cauchy for such equation. In \S 4 we 
obtain the longtime behaviour of the solutions of such a Cauchy problem, showing that the 
solutions break also in the longtime regime, and that such regime   
is essentially described by the longtime breaking formulae of the dKP solutions \cite{MS5},  
confirming the expected universal character of the dKP equation as prototype model in the description 
of the gradient catastrophe of two-dimensional waves. In \S 5 we characterize a class of RH spectral data 
allowing one to decouple and linearize the RH problem, generating a class of implicit solution of 2ddT 
parametrized by an arbitrary real function of one variable.

%%%%%%%%%%%%%%%%%%%%%%%%%%%%%%%%%%
\section{Nonlinear RH dressing}
%%%%%%%%%%%%%%%%%%%%%%%%%%%%%%%%%%

In this section we introduce the vector nonlinear RH problem enabling one 
to construct large classes of solutions of the 
Lax pair (\ref{Lax1}) and of the 2ddT equations (\ref{dT1}) and (\ref{dT3}).

\vskip 10pt
\noindent
{\it Proposition}. Consider the following vector nonlinear RH problem 
\beq\label{RH1}
\xi^+_j(\lambda)=\xi^-_j(\lambda)+R_j(\xi^-_1(\lambda)+\nu_1(\lambda),\xi^-_2(\lambda)+\nu_2(\lambda)),~~
\lambda\in \Gamma,~~j=1,2
\eeq  
on an arbitrary closed contour $\Gamma$ of the complex $\lambda$ plane, where 
$\vec R(\vec s)=(R_1(s_1,s_2),R_2(s_1,s_2))^T$ are given differentiable 
spectral data depending on the second argument $s_2$ through $exp(is_2)$ and satisfying the constraint
\beq\label{Ham_R}
\ba{l}
\{{\cal R}_1(s_1,s_2),{\cal R}_2(s_1,s_2)\}_{(s_1,s_2)}=1 , \\
{\cal R}_j(s_1,s_2):=s_j+R_j(s_1,s_2),~~~j=1,2,
\ea
\eeq
with
\beq
\{f,g\}_{(s_1,s_2)}:=f_{s_1}g_{s_2}-f_{s_2}g_{s_1};
\eeq
where $\nu_j,~j=1,2$ are the explicit functions
\beq\label{norm1}
\ba{l}
\vec\nu =\left(
\ba{c}
\nu_1 \\
\nu_2
\ea
\right)=\left(
\ba{c}
(\zeta_1\lambda +\zeta_2\lambda^{-1})v-t-\zeta_1\phi_{\zeta_1} \\
i\ln\lambda +i\frac{\phi_t}{2}
\ea
\right)
\ea
\eeq 
and $\vec\xi^{+}=(\xi^{+}_1,\xi^{+}_2)^T$ and $\vec\xi^{-}=(\xi^{-}_1,\xi^{-}_2)^T$ are the unknown vector solutions 
of the RH problem (\ref{RH1}), analytic rispectively inside and outside the contour $\Gamma$ and such that 
$\vec\xi^{-}\to \vec 0$ as $\lambda\to\infty$. Then, assuming that the above RH problem and its linearized form 
are uniquely solvable, we have the following results. 

\noindent
1) If 
\beq\label{closure1}
\ba{l}
\displaystyle\lim_{\lambda\to\infty}{(i\lambda \xi^-_2)}=\phi_{\zeta_1}e^{-\frac{\phi_t}{2}},     \\
i\xi^+_2(0)=\phi_t ,
\ea
\eeq
it follows that $\vec\pi^{\pm}=\vec\xi^{\pm}+\vec\nu$ are common eigenfunctions of $\hat L_{1,2}$: 
$\hat L_{1}\vec\pi^{\pm}=\hat L_{2}\vec\pi^{\pm}=\vec 0$ satisfying the relations
\beq
\{\pi^{\pm}_1,\pi^{\pm}_2\}_{(\lambda,t)}=
\lambda ({\pi^{\pm}_1}_{\lambda} {\pi^{\pm}_2}_t-{\pi^{\pm}_1}_t {\pi^{\pm}_1}_{\lambda})=i 
\eeq
and the potentials $\phi_{\zeta_1},\phi_t$, reconstructed through (\ref{closure1}), solve the 2ddT equation (\ref{dT1}). \\
2) In addition, if the variables $\zeta_1,\zeta_2$ are specified as in (\ref{elliptic}), if the RH data satisfy the 
additional reality constraint
\beq\label{reality_R}
\vec {\cal R}\left(\overline{\vec {\cal R}(\overline{\vec s})}\right)=\vec s ,~~~\forall \vec s\in\CC
\eeq
and $\Gamma$ is the unit circle, then the eigenfunctions satisfy the following symmetry 
relation:
\beq\label{symm}
\vec\pi^-(\lambda)=\overline{\vec\pi^+(1/\bar\lambda)}
\eeq
and $\phi\in\RR$.
 
\vskip 5pt
\noindent
{\bf Remark 1} The RH problem (\ref{RH1}) can obviously be formulated directly in terms of the eigenfunctions 
$\vec\pi^{\pm}$ 
as follows:
\beq\label{RH1b}
\pi^+_j(\lambda)={\cal R}_j(\pi^-_1(\lambda),\pi^-_2(\lambda))=
 \pi^-_j(\lambda)+R_j(\pi^-_1(\lambda),\pi^-_2(\lambda)),~~\lambda\in \Gamma , ~~j=1,2, 
\eeq 
with the normalization
\beq\label{norm1b} 
\vec\pi^-(\lambda)=\vec\nu(\lambda)+O(\lambda^{-1}),~|\lambda |>>1 .
\eeq
\vskip 5pt
\noindent
{\bf Remark 2} The dependence of $\vec R$ on $s_2$ through $exp(is_2)$ ensures that the solutions $\vec\xi^{\pm}$ 
of the RH problem do not exhibit the $\ln\lambda$ singularity contained in the normalization (\ref{norm1}). 
It follows that the eigenfunctions  
$\pi^{\pm}_2$ contain the $\ln\lambda$ singularity only as an additive singularity, while $\pi^{\pm}_1$ do not 
exhibit such singularity.  

\vskip 5pt
\noindent 
{\bf Remark 3} Before adding the closure conditions (\ref{closure1}), the solutions $\vec\xi^{\pm}$ of the 
RH problem depend, via the normalization (\ref{norm1}), on the undefined fields 
$\phi_t$ and $\phi_{\zeta_1}$, through the combination $(t+\zeta_1\phi_{\zeta_1})$; then the two closure conditions 
(\ref{closure1}) must be viewed as a nonlinear system 
of two algebraic equations for $\phi_t$ and $\phi_{\zeta_1}$ defining implicitly the solution $\phi_{\zeta_1},\phi_t$ 
of the 2ddT equation. Therefore, 
as in the dKP case \cite{MS5}, we expect that this spectral features be responsible for the wave 
breaking of localized initial data evolving according to the nonlinear wave equation (\ref{dT2}). 
Details on how two-dimensional waves evolving according to the 2ddT equation break will be presented 
elsewhere.  An alternative 
closure, perhaps useful in the reality reduction case described in part 2) of the above Proposition, is given by 
the equations
\beq\label{closure2}
\ba{l}
-i\frac{\xi^+_1(0)}{2}={\mbox Im }(t+z\phi_z),    \\
i\xi^+_2(0)=\phi_t .
\ea
\eeq
With this closure, indeed, we obtain a system of algebraic equations involving $t+z\phi_{z}$, its imaginary part  
and $\phi_t$.
 
\vskip 5pt
\noindent
{\bf Remark 4} The symmetry relations (\ref{symm}) are a distinguished example of the following symmetry of the 
common eigenfunctions of the Lax pair (\ref{Lax1}), when $\zeta_1=z,\zeta_2=\bar z$ as in (\ref{elliptic}) and 
$\phi\in\RR$:

\noindent   
{\it if $\psi(\lambda)$ is a solution of (\ref{Lax1}), then $\overline{\psi (1/\bar\lambda)}$ is a solution too}.
 
\vskip 5pt
\noindent
{\bf Remark 5} For part 2) of the above Proposition, when the contour $\Gamma$ is the unit circle, the 
RH problem is characterized by the following system of nonlinear integral equations for 
$\xi^{\pm}_j(\lambda),~|\lambda |=1$ (having parametrized $\lambda$ as in (\ref{param_lambda})):
\beq\label{int_equ}
\ba{l}
\xi^{\pm}_j(e^{-i\theta})-\frac{1}{2\pi}\int\limits_0^{2\pi}\frac{d\theta'}{1-(1\mp \epsilon)e^{i(\theta'-\theta)}}
R_j\Big((ze^{-i\theta'}+\bar z e^{i\theta'})v-t-z\phi_z+\xi^{-}_1(e^{-i\theta'}), \\
\theta'+i\frac{\phi_t}{2}+\xi^{-}_2(e^{-i\theta'})\Big)=0,~~~j=1,2,
\ea
\eeq 
and the closure conditions (\ref{closure1}) read
\beq\label{sol1a}
\ba{l}
\phi_ze^{-\frac{\phi_t}{2}}=\frac{1}{2\pi i}\int\limits_0^{2\pi}d\theta e^{-i\theta} 
R_2\Big((ze^{-i\theta}+\bar z e^{i\theta})v-t-z\phi_z+\xi^{-}_1(e^{-i\theta}), \\
\theta+i\frac{\phi_t}{2}+\xi^{-}_2(e^{-i\theta})\Big)=0,
\ea
\eeq
\beq\label{sol1b}
\ba{l}
\phi_t=-\frac{1}{2\pi i}\int\limits_0^{2\pi}d\theta 
R_2\Big((ze^{-i\theta}+\bar z e^{i\theta})v-t-z\phi_z+\xi^{-}_1(e^{-i\theta}), 
\theta+ \\ i\frac{\phi_t}{2}+ \xi^{-}_2(e^{-i\theta})\Big)=0.
\ea
\eeq

\vskip 10pt
\noindent
{\it Proof}. For part 1), we first apply the operators $\hat L_{1,2}$ to the RH problem (\ref{RH1b}), obtaining the 
linearized RH problem $\hat L_j\vec\pi^+=A\hat L_j\vec\pi^-$, where $A$ is the Jacobian matrix 
of the transformation 
(\ref{RH1b}): $A_{ij}=\partial {\cal R}_i/\partial s_j,~i,j=1,2$. Since, due to the normalization (\ref{norm1b}), 
$\hat L_j\vec\pi^-\to \vec 0$ as $\lambda\to\infty$, it follows that, by uniqueness, $\vec\pi^{\pm}$ are common 
eigenfunctions of the vector fields $\hat L_{1,2}$: $\hat L_{1,2}\vec\pi^{\pm}=\vec 0$ and, consequently, that $\phi_t,\phi_z$ 
are solutions of the 2ddT equation (\ref{dT1}). Then the eigenfunctions exhibit the following asymptotics: 
\beq\label{asympt_lambda_pi}
\ba{l}
\pi^-_1=\lambda z v-t-z\phi_z +\lambda^{-1} \left(\bar z v +a^- v^{-1}\right)+O(\lambda^{-2}),~~
|\lambda|>>1, \\
\pi^+_1=\lambda^{-1}\bar z v-t-\bar z\phi_{\bar z}+\lambda\left(z v+a^+ v^{-1}\right)+ 
O(\lambda^{2}),~~|\lambda|< <1, \\
\pi^{-}_2= i\ln \lambda+i\frac{\phi_t}{2}-i\lambda^{-1}\phi_z v^{-1}+O(\lambda^{-2}),~~|\lambda|>>1, \\
\pi^{+}_2= i\ln \lambda-i\frac{\phi_t}{2}+i\lambda \phi_{\bar z}v^{-1}+O(\lambda^2),~~|\lambda|< < 1 ,
\ea
\eeq    
where
\beq\label{a1}
a^-_{\bar z}=-a^+_{z}=(ze^{\phi_t})_z-(\bar z e^{\phi_t})_{\bar z}.
\eeq  
implying the closure conditions (\ref{closure1}),(\ref{closure2}). Since, from (\ref{RH1b}), 
$\{\pi^+_1,\pi^+_2 \}_{(t,\lambda)}=\{{\cal R}_1,{\cal R}_2\}_{\vec\pi^-}\{\pi^-_1,\pi^-_2 \}_{(t,\lambda)}$, $\lambda\in\Gamma$,  
equation (\ref{Ham_R}) implies that $\{\pi^+_1,\pi^+_2 \}_{(t,\lambda)}=\{\pi^-_1,\pi^-_2 \}_{(t,\lambda)},~\lambda\in\Gamma$; i.e., 
the Poisson brackets of the $\pm$ eigenfunctions are analytic in the whole complex $\lambda$ plane.  
Since $\{\pi^-_1,\pi^-_2 \}\to i$ as $\lambda\to\infty$, it follows that 
$\{\pi^+_1,\pi^+_2 \}=\{\pi^-_1,\pi^-_2 \}=i$. For part 2), applying $\vec{\cal R}(\cdot )$ to the complex conjugate of 
the RH problem (\ref{RH1b}) and using the reality condition (\ref{reality_R}) it follows that 
$\overline{\vec\pi^+(\lambda )}=\vec\pi^-(\lambda ),~|\lambda |=1$. By the Schwartz reflection principle, it follows 
the symmetry (\ref{symm}) and, using the Lax pair (\ref{Lax1}), the reality condition $\phi\in\RR$. $\Box$

%%%%%%%%%%%%%%%%%%%%%%%%%%%%%%%%%%%%%%%%%%%%%%%%%%%%%%%%%%%%
\section{The Cauchy problem for the 2ddT equation}
%%%%%%%%%%%%%%%%%%%%%%%%%%%%%%%%%%%%%%%%%%%%%%%%%%%%%

In this section we present the formal solution of the Cauchy problem for the wave form of the 2ddT equation:
\beq\label{Cauchy_data}
\ba{l}
\left(e^{\phi_t}\right)_{t}=\phi_{xx}+\phi_{yy}, ~~~x,y\in\RR,~~t>0,~~\phi(x,y,t)\in\RR , \\
\phi(x,y,0)=A(x,y),~~\phi_t(x,y,0)=B(x,y).
\ea
\eeq
where the assigned initial conditions $A(x,y),B(x,y)$   
are localized in the ($x,y$) plane for $x^2+y^2 \to\infty$. To 
do it, we use the IST for vector fields developed in \cite{MS1,MS2,MS3,MS4}. 

In this respect, we recall two basic facts: since the 
Lax pair of the 2ddT is made of vector fields, i) {\it the space of eigenfunctions is a ring} 
(if $f_1$ and $f_2$ are eigenfunctions, any differentiable function $F(f_1,f_2)$ is an eigenfunction);
ii) since the vector fields are also Hamiltonian, 
{\it the space of eigenfunctions is also a Lie algebra, whose Lie bracket is the Poisson bracket 
(\ref{PB1})} (if $f_1$ and $f_2$ are eigenfunctions, also $\{f_1,f_2\}_{(\lambda,t)}$ is an eigenfunction). 

Multiplying the first and second equations of (\ref{Lax1}) (with 
$\zeta_1=z,~\zeta_2=\bar z$ as in (\ref{elliptic})) by $\lambda^{-1}$ and $\lambda$ 
respectively, then adding and subtracting the resulting equations, one obtains the equivalent and more convenient Lax pair:
\beq\label{Lax2a}
\ba{l}
\hat{\cal L}_1\psi:=\lambda\psi_{\bar z}-\lambda^{-1}\psi_{z}-\Big(-2v_t+\lambda\frac{\phi_{\bar zt}}{2}+
\lambda^{-1}\frac{\phi_{zt}}{2}\Big)\lambda\psi_{\lambda}=0, 
\ea
\eeq
\beq\label{Lax2b}
\ba{l}
\hat{\cal L}_2\psi:=\psi_t+\frac{v^{-1}}{2}\left(\lambda\psi_{\bar z}+\lambda^{-1}\psi_z\right)-
\frac{v^{-1}}{4}\left(\lambda\phi_{\bar z t}-\lambda^{-1}\phi_{zt}\right)\lambda\psi_{\lambda}=0,
\ea
\eeq
where the first equation must be viewed as the spectral problem (in which $v_t$ shall be replaced, in the direct 
problem, 
by $(\phi_{z\bar z}/2)exp(-\phi_t/2)$, due to (\ref{dT3})) and the second equation as $t$-evolution of the eigenfunction.

%%%%%%%%%%%%%%%%%%%%%%%%%%%%%%%%%%%%%%%
\vskip 10pt
\noindent
{\bf Eigenfunctions and spectral data}. Now we introduce the Jost and analytic eigenfunctions 
for the spectral problem (\ref{Lax2a}). Since the  
associated undressed operator: $\lambda\partial_{\bar z}-\lambda^{-1}\partial_z$ coincides with 
the undressed operator of the spectral problem for the ($2+1$)-dimensional self-dual Yang-Mills equation 
\cite{MZ}, the construction of the Jost and analytic Green's functions is taken from there.
   
We define Jost eigenfunctions of the spectral problem (\ref{Lax2a}) on the unit circle of the complex 
$\lambda$ plane, using the parametrization (\ref{param_lambda}). Introducing the convenient real variables 
$\xi,\eta,\theta'$ as follows:
\beq\label{change}
\ba{l}
\xi=\cos\theta ~x+\sin\theta ~y, \\
\eta=-\sin\theta ~x+\cos\theta ~y, \\
\theta'=\theta,
\ea
\eeq
the Lax pair (\ref{Lax2a}),(\ref{Lax2b}) becomes
\beq\label{Lax3a}
\ba{l}
\hat{\cal L}_1\psi:=\psi_{\eta}-\frac{1}{2}\left[-\left(\phi_{\xi\xi}+\phi_{\eta\eta}\right)v^{-1}+\phi_{\xi t}\right]
(\eta\psi_{\xi}-\xi\psi_{\eta}+\psi_{\theta'})=0,                                  
\ea
\eeq 
\beq\label{Lax3b}
\ba{l}
\hat{\cal L}_2\psi:=\psi_{t}+v^{-1}\psi_{\xi}-(v^{-1})_{\eta}(\eta\psi_{\xi}-\xi\psi_{\eta}+\psi_{\theta'})=0.
\ea
\eeq 
A convenient basis of Jost eigenfunctions are the solutions $f_1$ and $f_2$ of equation 
(\ref{Lax3a}) satisfying the boundary conditions
\beq
\vec f(\xi,\eta,\theta'):=\left(
\ba{c}
f_1(\xi,\eta,\theta') \\
f_2(\xi,\eta,\theta')
\ea
\right) \to 
\left(
\ba{c}
\xi \\
\theta'
\ea
\right),~~\mbox{as }\eta\to -\infty ;
\eeq
they are characterized by the linear integral equation
\beq
\ba{l}
\vec f =\left(
\ba{c}
\xi \\
\theta'
\ea
\right)+
\frac{1}{2}\int\limits_{-\infty}^{\eta}d\eta'\left[-(\phi_{\xi\xi}+\phi_{\eta'\eta'})v^{-1}+\phi_{\xi t}\right]
(\eta'\vec f_{\xi}-\xi\vec f_{\eta'}+\vec f_{\theta'}).
\ea
\eeq
It follows that $f_1(\xi,\eta,\theta)$ and $f_2(\xi,\eta,\theta)-\theta$ are $2\pi$-periodic in $\theta$.

The $\eta\to \infty$ limit of $\vec f$ defines the 
scattering vector $\vec\sigma (\xi, \theta) = (\sigma_1(\xi, \theta), \sigma_2 (\xi, \theta))^T$ as follows 
\beq
\vec f(\xi,\eta,\theta) \to \vec {\cal S}(\xi,\theta)=
\left(
\ba{c}
\xi \\
\theta
\ea
\right)+
\vec\sigma(\xi,\theta),~~\mbox{as }~~\eta\to \infty ;
\eeq
namely:
\beq
\vec\sigma(\xi,\theta)=\frac{1}{2}\int\limits_{\RR}d\eta\left[-(\phi_{\xi\xi}+\phi_{\eta\eta})v^{-1}+\phi_{\xi t}\right]
(\eta\vec f_{\xi}-\xi\vec f_{\eta}+\vec f_{\theta}).
\eeq
Also the scattering vector is $2\pi$-periodic in $\theta$: 
$\vec\sigma(\xi,\theta+2\pi)=\vec\sigma(\xi,\theta)$; i.e., its dependence on the second 
argument $\theta$ is through $exp(i\theta)$.  

The analytic eigenfunctions of (\ref{Lax2a}) are defined instead via the integral equations:
\beq\label{int-equ-anal}
\ba{l}
\vec\psi^{\pm}(z,\bar z,\lambda)=\left(
\ba{c}
\psi^{\pm}_1(z,\bar z,\lambda) \\
\psi^{\pm}_2(z,\bar z,\lambda)
\ea
\right)=\left(
\ba{c}
\lambda z+\lambda^{-1}\bar z \\
i\ln\lambda
\ea
\right)+          \\ 
\frac{i}{4}\int_{\CC}dz'\wedge d\bar z'G^{\pm}(z-z',\bar z-\bar z',\lambda)
\left[-\phi_{z\bar z}v^{-1}+\lambda\frac{\phi_{\bar z' t}}{2}+\lambda^{-1}\frac{\phi_{z't}}{2}\right]
\lambda {\vec\psi^{\pm}}_{\lambda}(z',\bar z',\lambda), 
\ea
\eeq
where $G^{\pm}$ are the analytic Green's functions
\beq\label{Green1}
G^{\pm}(z,\bar z,\lambda)=\mp\frac{1}{\pi}\frac{1}{\left(\lambda z+\lambda^{-1}\bar z\right)},~~
\mbox{sgn}(1-|\lambda |)=\pm 1
\eeq 
such that $\lambda G^{\pm}_{\bar z}-\lambda^{-1}G^{\pm}_z=\delta (z)$, reducing, on the unit circle $|\lambda|=1$, 
to
\beq\label{Green2}
G^{\pm}(z,\bar z,\lambda)=\mp \frac{1}{\pi}\frac{1}{\xi\mp i\epsilon \eta}, ~~ 0< \epsilon < < 1 .
\eeq

Since $G^+$ and $G^-$ are analytic 
respectively inside and outside the unit circle of the complex $\lambda$ plane, then  
$(\psi^+_1,\psi^+_2)$  and $(\psi^-_1,\psi^-_2)$ are also analytic, respectively, inside and outside 
the unit circle of the complex $\lambda$ plane, after subtracting their singular parts, given respectively  
by $(\lambda^{-1}\bar z v,i\ln \lambda)$ and $(\lambda z v,i\ln \lambda)$, 
as it can be seen by solving the integral equations (\ref{int-equ-anal}) by iteration or from the following  
$\lambda$ - asymptotics:
\beq\label{asympt_lambda_psi}
\ba{l}
\psi^-_1=\lambda z v- z\phi_z +\lambda^{-1} \left(\bar z v +a^- v^{-1}\right)+O(\lambda^{-2}),~~
|\lambda|>>1, \\
\psi^+_1=\lambda^{-1}\bar z v-\bar z\phi_{\bar z}+\lambda\left(z v+a^+ v^{-1}\right)+ 
O(\lambda^{2}),~~|\lambda|< <1, \\
\psi^{-}_2= i\ln \lambda+i\frac{\phi_t}{2}-i\lambda^{-1}\phi_z v^{-1}+O(\lambda^{-2}),~~|\lambda|>>1, \\
\psi^{+}_2= i\ln \lambda-i\frac{\phi_t}{2}+i\lambda \phi_{\bar z}v^{-1}+O(\lambda^2),~~|\lambda|< < 1 ,
\ea
\eeq
where $a^{\pm}$ are defined in (\ref{a1}).

In addition, equations (\ref{Green2}) imply the limits
\beq
\ba{l}
G^+(z-z',\bar z-\bar z',\lambda)\to -\frac{1}{\pi}\frac{1}{\xi-\xi'\pm i\varepsilon},~~\mbox{as } \eta\to\mp\infty, \\
G^-(z-z',\bar z-\bar z',\lambda)\to \frac{1}{\pi}\frac{1}{\xi-\xi'\mp i\varepsilon},~~\mbox{as } \eta\to\mp\infty.
\ea
\eeq
Therefore, on the unit circle $|\lambda|=1$, the $\eta\to -\infty$ limit of $(\psi^+_1,\psi^+_2)$  
and $(\psi^-_1,\psi^-_2)$ are 
analytic respectively in the upper and lower parts of the complex $\xi$ plane, while the $\eta\to \infty$ limit 
of $(\psi^+_1,\psi^+_2)$  and $(\psi^-_1,\psi^-_2)$ are 
analytic respectively in the lower and upper parts of the complex $\xi$ plane. This mechanism, first 
observed in \cite{MZ}, plays an important role in the IST for vector fields 
(see \cite{MS1},\cite{MS2},\cite{MS3},\cite{MS4}). 

Since the Jost eigenfunctions $\vec f=(f_1,f_2)^T$ are a good basis in the space of eigenfunctions 
of the spectral problem (\ref{Lax3a}) for $|\lambda |=1$, one can express the analytic eigenfunctions in terms 
of them through the following formulae, valid for $|\lambda |=1$:
\beq\label{basis-equ}
\vec\psi^{\pm}=\vec{\cal K}^{\pm}(\vec f)=\vec f +\vec\chi^{\pm}(f_1,f_2),
\eeq
defining the spectral data $\vec\chi^{\pm}$ as differentiable functions of two arguments. 
In the $\eta\to -\infty$ limit, equations (\ref{basis-equ}) reduce to
\beq\label{basis-equ-infty}
\ba{l}
\displaystyle\lim_{\eta\to -\infty}{\vec\psi^{\pm}} -
\left(
\ba{c}
\xi \\
\theta
\ea
\right)=\vec\chi^{\pm}(\xi,\theta),
\ea
\eeq  
implying that i) $\vec\chi^+(\xi,\theta)$ and $\vec\chi^-(\xi,\theta)$ are analytic in the first variable 
$\xi$ respectively in the upper and lower half parts of the complex $\xi$ plane, and ii) $\vec\chi^{\pm}(\xi,\theta)$  
are $2\pi$-periodic in $\theta$: $\vec\chi^{\pm}(\xi,\theta+2\pi)=\vec\chi^{\pm}(\xi,\theta)$ (their dependence on the 
second argument $\theta$ is through $exp(i\theta)$).  

At $\eta\to\infty$, equations (\ref{basis-equ}) reduce to 
\beq\label{basis-equ+infty}
\ba{l}
\displaystyle\lim_{\eta\to \infty}{\vec\psi^{\pm}} -\left(
\ba{c}
\xi \\
\theta
\ea
\right)=
\vec \sigma +\vec\chi^{\pm}(\xi+\sigma_1,\theta+\sigma_2).
\ea
\eeq  
Applying the operator $\int_{\RR}d\xi \int\limits_{0}^{2\pi}\frac{d\theta}{2\pi}e^{-i(\omega\xi+n\theta)}\cdot,~n\in\ZZ$ 
to equations (\ref{basis-equ+infty}) and using the 
above established analiticity properties in $\xi$ and the $2\pi$-periodicity in $\theta$, we obtain the following linear 
integral equations connecting the (Fourier transforms of the) scattering data $\vec\sigma$ to the 
(Fourier transforms of the) spectral data $\vec\chi^{\pm}$:
\beq\label{sigma-chi}
\ba{l}
{\tilde{\vec\chi}^{\pm}}(\omega,n)+H(\pm\omega)\left(
\tilde{\vec\sigma}(\omega,n)+
\int\limits_{\RR}d\omega'\sum\limits_{n'=-\infty}^{\infty}{\tilde{\vec\chi}^{\pm}}(\omega',n')Q(\omega',n',\omega,n)\right)=
\vec 0 ,
\ea
\eeq 
where $H$ is the Heaviside step function and 
\beq
\ba{l}
Q(\omega',n',\omega,n)=
\int\limits_{\RR}\frac{d\xi}{2\pi}\int\limits_{0}^{2\pi}\frac{d\theta}{2\pi}e^{i(\xi(\omega'-\omega)+(n'-n)\theta)}
\left(e^{i(\omega'\sigma_1(\xi,\theta)+n'\sigma_2(\xi,\theta))}-1\right),                   \\
{\tilde{\vec\chi}^{\pm}}(\omega,n)=\int\limits_{\RR}d\xi\int\limits_{0}^{2\pi}\frac{d\theta}{2\pi}
e^{-i(\omega\xi+n\theta)}{\vec\chi}^+(\xi,\theta),                                          \\
{\tilde{\vec\sigma}}(\omega,n)=\int\limits_{\RR}d\xi\int\limits_{0}^{2\pi}\frac{d\theta}{2\pi}
e^{-i(\omega\xi+n\theta)}{\vec\sigma}^+(\xi,\theta).
\ea
\eeq

At last, eliminating, from equations (\ref{basis-equ}), the Jost eigenfunctions $\vec f$, one obtains, through 
algebraic manipulation, the following vector nonlinear RH problem on the unit circle of the complex  
$\lambda$ plane:
\beq\label{RH2}
\ba{l}
\psi^+_1={\cal R}_1(\psi^-_1,\psi^-_2)=\psi^-_1+R_1(\psi^-_1,\psi^-_2),~~~~|\lambda |=1,   \\
\psi^+_2={\cal R}_2(\psi^-_1,\psi^-_2)=\psi^-_2+R_2(\psi^-_1,\psi^-_2).
\ea
\eeq
We remark that the $2\pi$-periodicity properties of the scattering data $\vec\chi^{\pm}(\xi,\theta)$ in 
the variable $\theta$ imply 
that the dependence of $\vec R$ on the second argument $s_2$ is also through $exp(is_2)$, to guaranty that the  
$\ln\lambda$ singularity is just an additive one for $\psi^{\pm}_2$, and is absent for $\psi^{\pm}_1$.  

Recapitulating, in the direct problem, at $t=0$, we go from the initial conditions $\phi,\phi_t$  
of the 2ddT equation to the initial scattering vector $\vec\sigma(\xi,\theta)$; 
from it we construct, through the linear integral equations (\ref{sigma-chi}), the scattering data 
$\vec\chi^{\pm}(\xi,\theta)$ and, through algebraic manipulation, the RH spectral data 
$\vec R(\vec s)=(R_1(s_1,s_2),R_2(s_1,s_2))$. In the   
inverse problem, one gives the RH spectral data $\vec R(\vec s)$ and reconstructs the vector solutions 
$\vec\psi^{\pm}$ of the RH problem (\ref{RH2}), defined by the normalization:
\beq\label{norm2}
\vec\psi^-=
\left(
\ba{c}
(z \lambda +\bar z\lambda^{-1})e^{\frac{\phi_t}{2}}- z\phi_{z} \\
i\ln\lambda +i\frac{\phi_t}{2}
\ea
\right)+\vec O(\lambda^{-1}),~~|\lambda |>>1 .
\eeq
At last, the closure conditions 
\beq\label{closure3}
\ba{l}
\displaystyle\lim_{\lambda\to\infty}{\lambda(i\psi^-_2+\ln\lambda)}=\phi_{z}e^{-\frac{\phi_t}{2}},   \\
\displaystyle\lim_{\lambda\to 0}{(i\psi^+_2+\ln\lambda ) }=\frac{\phi_t}{2},
\ea
\eeq
consequences of the asymptotics (\ref{asympt_lambda_psi}), 
allow one to reconstruct the solution of the 2ddT equation through the solution of a system of two algebraic 
equations for $\phi_t$ and $\phi_z$.

%%%%%%%%%%%%%%%%%%%%%%%%%%%%%%%%%%%%%%%%%%%%%%
\vskip 10pt
\noindent
{\bf Time evolution of the spectral data}. To construct the $t$-evolution of the spectral data we observe that 
$\vec f$ and $\vec\psi^{\pm}$, eigenfunctions of the spectral problem (\ref{Lax2a}): 
$\hat{\cal L}_1\vec f=\hat{\cal L}_1\vec\psi^{\pm}=\vec 0$, are solutions of the following equations involving the 
second Lax operator $\hat{\cal L}_2\vec f=\hat{\cal L}_2\vec\psi^{\pm}=(1,0)^T$, implying the following elementary  
time evolutions of the data:
\beq
\ba{l}
\vec\sigma(\xi,\theta,t)=\vec\sigma(\xi-t,\theta,0),~~\vec\chi^{\pm}(\xi,\theta,t)=\vec\chi^{\pm}(\xi-t,\theta,0),~~ \\
\vec R(\xi,\theta,t)=\vec R(\xi-t,\theta,0). 
\ea
\eeq
In addition, it follows that the common Jost eigenfunctions $\vec J$ and the common  
analytic eigenfunctions $\vec\pi^{\pm}$ of the Lax pair (\ref{Lax2a}),(\ref{Lax2b}) are obtained from $\vec f$ and 
$\vec\psi^{\pm}$ simply as follows:
\beq\label{common-eigen}
\ba{l}
\vec J:=\vec f-t(1,0)^T,                  \\
\vec\pi^{\pm}:=\vec\psi^{\pm}-t(1,0)^T .
\ea
\eeq
It is easy to verify that the analytic eigenfunctions $\vec\pi^{\pm}$, the RH data $\vec R(\vec s)$ and the 
associated RH problem of this section coincide with those appearing in the dressing construction of \S 2.

%%%%%%%%%%%%%%%%%%%%%%%%%%%%%%%%%%%%%%%%%%%%%%%%
\vskip 10pt
\noindent
{\bf Hamiltonian constraints on the data}. The Hamiltonian character of the 2ddT dynamics implies the following 
formulae for the Poisson brackets of the relevant eigenfunctions:
\beq\label{PB_J}
\{J_1,J_2\}_{(\lambda,t)}=\{\pi^{\pm}_1,\pi^{\pm}_2\}_{(\lambda,t)}=i
\eeq
which, in turn, imply that the transformations $\vec s\to \vec{\cal K}^{\pm}(\vec s)$ and $\vec s\to \vec{\cal R}(\vec s)$ 
are canonical: 
\beq\label{PB_K} 
\{{\cal K}^{\pm}_1,{\cal K}^{\pm}_2\}_{(s_1,s_2)}=\{{\cal R}_1,{\cal R}_2\}_{(s_1,s_2)}=1.
\eeq
To prove (\ref{PB_J}), one first shows that $J_3:=\{J_1,J_2\}_{(\lambda,t)}\to i$ as $\eta\to -\infty$,  
$\pi^{-}_3:=\{\pi^{-}_1,\pi^{-}_2\}_{(\lambda,t)}\to i$ as $\lambda \to \infty$, 
$\pi^{+}_3:=\{\pi^{+}_1,\pi^{+}_2\}_{(\lambda,t)}\to i$ as $\lambda \to 0$. Since the vector fields are 
Hamiltonian, $J_3,\pi^{\pm}_3$ are also common eigenfunctions, and equations (\ref{PB_J}) hold, by uniqueness. Equations 
(\ref{PB_K}) are consequences of (\ref{PB_J}) and of the relations
\beq
\vec\pi^{\pm}=\vec{\cal K}^{\pm}\left(\vec J\right),~~~~\vec\pi^+=\vec{\cal R}\left(\vec\pi^-\right).
\eeq 

%%%%%%%%%%%%%%%%%%%%%%%%%%%%%%%%%%%%%%%%%%%%%%%%
\vskip 10pt
\noindent
{\bf Reality constraints}. The definition (\ref{common-eigen}b) and the condition $\phi\in\RR$ imply the symmetry 
relations 
\beq
\vec\pi^-(\lambda)=\overline{\vec\pi^+(1/\bar\lambda)};
\eeq
consequently, from (\ref{RH1}), the reality constraint (\ref{reality_R}) on the RH data holds true.

%%%%%%%%%%%%%%%%%%%%%%%%%%%%%%%%%%%%%%%%%%%%%%%%
\vskip 10pt
\noindent
{\bf Small field limit and Radon Transform}. As for the IST of the heavenly \cite{MS2} and dKP \cite{MS3} equations,  
in the small field limit $|\phi |,|\phi_t|<<1$, the direct and inverse 
spectral transforms presented in this section reduce to the direct and inverse Radon transform \cite{Radon}. Indeed, 
the mapping from the initial data $\{A(x,y),B(x,y) \}$ to the scattering vector $\vec\sigma$ reduces 
to the direct Radon transform:
\beq\label{SFL_DP}
\ba{l}
\vec\sigma(\xi,\theta) \sim \frac{1}{2}\int\limits_{\RR}\left(
\ba{c}
\eta \\
1
\ea
\right)\Big[-(\partial^2_{\xi}+\partial^2_{\eta})A(x(\xi,\eta,\theta),y(\xi,\eta,\theta))+  \\
\partial_{\xi}B(x(\xi,\eta,\theta),y(\xi,\eta,\theta))\Big]d\eta, \\
x(\xi,\eta,\theta)=\xi\cos\theta -\eta\sin\theta ,~~~y(\xi,\eta,\theta)=\xi\sin\theta +\eta\cos\theta ,
\ea
\eeq 
while the spectral data $\vec\chi^{\pm}$ and $\vec R$ are constructed from $\vec\sigma$ as follows:
\beq\label{sigma_chi_R}
\vec\chi^{\pm}(\xi,\theta)\sim -\hat P^{\pm}_{\xi}\vec\sigma(\xi,\theta),~~
\vec R(\xi,\theta)\sim -i\hat{\cal H}_{\xi}\vec\sigma(\xi,\theta),
\eeq
where $\hat P^{\pm}_{\xi}$ and $\hat{\cal H}_{\xi}$ are rispectively the $(\pm)$ analyticity projectors and 
the Hilbert transform in the variable $\xi$:
\beq
\hat P^{\pm}_{\xi}g(\xi):=\pm \frac{1}{2\pi i}\int_{\RR}\frac{d\xi'}{\xi'-(\xi\pm i0)}g(\xi'),~~
\hat{\cal H}_{\xi}g(\xi):=\frac{1}{\pi}P\int\limits_{\RR}\frac{d\xi'}{\xi-\xi'}g(\xi').
\eeq
At last, the first of the closure conditions (\ref{closure3}) of the inverse problem reduces to 
the inverse Radon transform
\beq\label{SFL_IP}
\ba{l}
\phi_t(x,y,t)\sim -\frac{1}{2\pi i}\int\limits_0^{2\pi}d\theta R_2(\xi-t,\theta)\sim 
-\frac{1}{2\pi^2}\int\limits_0^{2\pi}d\theta ~P\int\limits_{\RR}\frac{d\xi'}{\xi'-(\xi-t)}\sigma_2(\xi',\theta), \\
\xi=x\cos\theta+y\sin\theta ,
\ea
\eeq
that can be shown to be equivalent to the well-known Poisson formula
\beq
\ba{l}
\phi(x,y,t)=\partial_t\int_{\RR^2}\frac{dx'dy'}{2\pi}L(x-x',y-y',t)A(x',y')+ \int_{\RR^2}\frac{dx'dy'}{2\pi}L(x-x', \\ 
y-y',t)B(x',y'),
\ea
\eeq
where 
\beq
L(x,y,t):=\frac{H(t^2-x^2-y^2)}{\sqrt{t^2-x^2-y^2}}
\eeq
and $H(\cdot)$ is the Heaviside step function, describing the solution of the Cauchy problem
\beq\label{Cauchy_data_lin}
\ba{l}
\phi_{tt}=\phi_{xx}+\phi_{yy}, ~~~x,y\in\RR,~~t>0,~~\phi(x,y,t)\in\RR , \\
\phi(x,y,0)=A(x,y),~~\phi_t(x,y,0)=B(x,y).
\ea
\eeq
for the linear wave equation in 2+1 dimensions.

%%%%%%%%%%%%%%%%%%%%%%%%%%%%%%%%%%%%%%%%%%%%%%%%%%%%%%%%%%%%
\section{The longtime behaviour of the solutions}
%%%%%%%%%%%%%%%%%%%%%%%%%%%%%%%%%%%%%%%%%%%%%%%%%%%%%

In this section we show, as it was done in the dKP case \cite{MS5}, that the spectral mechanism causing the breaking 
of a localized initial condition evolving according to the 2ddT equation is present also in the longtime 
regime. We will actually show that {\it the longtime breaking of the 2ddT solutions 
is essentially described by the longtime breaking formulae of the dKP solutions} found in \cite{MS5}; 
this is an important confirmation of the expected  
{\it universal character of the dKP equation as prototype model in the description of the gradient catastrophe of  
two-dimensional waves}. 

We remark that it is clearly meaningful to study the longtime behaviour of the solutions of the 2ddT equation 
only if no breaking takes place before, at finite time. In this section we assume that the initial condition be  
small, then the nonlinearity becomes important only in the longtime regime and no breaking takes place before.  

Motivated by the longtime behaviour  
of the solutions of the linear wave equation $u_{tt}=u_{xx}+u_{yy}$, localized, with amplitude $O(t^{-\frac{1}{2}})$, in 
the region $\sqrt{x^2+y^2}-t=O(1)$, we study the longtime behaviour 
of the solutions of the 2ddT equation in the space-time region
\beq\label{asympt_region}
\ba{l}
z=\frac{t+r}{2}e^{i\alpha},~~~~~~~\alpha,r\in\RR ,~~\alpha=O(1),~~t>>1,
\ea
\eeq
implying that
\beq
r=\sqrt{x^2+y^2}-t,~~~~\alpha=\arctan \frac{y}{x}.
\eeq
Substituting (\ref{asympt_region}) into the integral equations (\ref{int_equ}) and keeping in mind that, 
in the longtime regime, $\phi_t$ is small, so that, f. i., $v\sim 1+\phi_t/2+\phi^2_t/8$, we obtain
\beq\label{as_int_equ1}
\ba{l}
\xi^{\pm}_j(\lambda)-\frac{1}{2\pi}\int\limits_0^{2\pi}\frac{d\theta'}{1-(1\mp \epsilon)e^{i(\theta'-\theta)}}
R_j\Big(-2t\sin^2\left(\frac{\theta'-\alpha}{2}\right)+r\cos(\theta'-\alpha)+   \\
\frac{t+r}{2}\cos(\theta'-\alpha)\phi_t(1+\frac{\phi_t}{4})-z\phi_z+\xi^{-}_1(e^{-i\theta'}),                     
\theta'+\xi^{-}_2(e^{-i\theta'})\Big) \sim 0,~~j=1,2.
\ea
\eeq
Since the main contribution to these integrals occurs when $\sin((\theta'-\alpha)/2)\sim 0$, we make the 
change of variable $\theta'=\alpha-\mu'/\sqrt{t}$, obtaining
\beq\label{as_int_equ2}
\ba{l}
\xi^{\pm}_j(\lambda)-\frac{1}{2\pi\sqrt{t}}
\int\limits_{\RR}\frac{d\mu'}{1-(1\mp \epsilon)e^{i(\alpha-\theta-\frac{\mu'}{\sqrt{t}})}}
R_j\Big( -\frac{{\mu'}^2}{2}+ X + 
 \xi^{-}_1\left(e^{-i(\alpha-\frac{\mu'}{\sqrt{t}}) }\right),  \\
\alpha + \xi^{-}_2\left(e^{-i(\alpha-\frac{\mu'}{\sqrt{t}})}\right)\Big) \sim 0,~~~~~~~~j=1,2,
\ea
\eeq
where
\beq
X:=r+\frac{t+r}{2}\phi_t+\frac{t}{8}\phi^2_t-z\phi_z .
\eeq
If $|\theta-\alpha|>>t^{-1/2}$, equations (\ref{as_int_equ2}) imply that $\xi^{\pm}_j(\lambda)=O(t^{-1/2})$:
\beq\label{as_int_equ3}
\ba{l}
\xi^{\pm}_j(\lambda)\sim \frac{1}{2\pi\sqrt{t}\left(1-(1\mp \epsilon)e^{i(\alpha-\theta)}\right)}
\int\limits_{\RR}d\mu'R_j\Big(-\frac{{\mu'}^2}{2}+X +
\xi^{-}_1\left(e^{-i(\alpha-\frac{\mu'}{\sqrt{t}})}\right), \\
\alpha + \xi^{-}_2\left(e^{-i(\alpha-\frac{\mu'}{\sqrt{t}})}\right)\Big),~~j=1,2.
\ea
\eeq
If, instead, $\theta-\alpha=-\mu t^{-1/2}$, $|\mu |=O(1)$, then $\xi^{\pm}_j(\lambda)=O(1)$:
\beq\label{as_int_equ4}
\ba{l}
\xi^{\pm}_j\left(e^{-i(\alpha-\frac{\mu}{\sqrt{t}})}\right)\sim \frac{1}{2\pi i}
\int\limits_{\RR}\frac{d\mu'}{\mu'-(\mu \pm i\epsilon)}
R_j\Big(-\frac{{\mu'}^2}{2}+X+ 
\xi^{-}_1\left(e^{-i(\alpha-\frac{\mu'}{\sqrt{t}})}\right), \\
\alpha + \xi^{-}_2\left(e^{-i(\alpha-\frac{\mu'}{\sqrt{t}})}\right)\Big),~~j=1,2.
\ea
\eeq
Therefore it is not possible to neglect, in the above integral equations, $\xi^{-}_j,~j=1,2$ in the arguments of 
$R_j,~j=1,2$; it follows that these integral equations remain nonlinear even in the longtime regime.

At last, using equations (\ref{as_int_equ3}), the asymptotic form of the closure conditions read, for $t>>1$:
\beq\label{as_phi_t}
\ba{l}
\phi_t\sim -\frac{1}{2\pi i\sqrt{t}}\int\limits_{\RR}d\mu'R_2
\Big(-\frac{{\mu'}^2}{2}+X + \xi^{-}_1\left(e^{-i(\alpha-\frac{\mu'}{\sqrt{t}})}\right), \\
\alpha + \xi^{-}_2\left(e^{-i(\alpha-\frac{\mu'}{\sqrt{t}})}\right)\Big),         \\
\ea
\eeq 
\beq\label{as_phi_z}
\ba{l}
\phi_z=-e^{-i\alpha}\phi_t(1+\frac{\phi_t}{2})\left(1+O(t^{-1}) \right).
\ea
\eeq 
Comparing (\ref{as_phi_t}) and (\ref{as_phi_z}), and using (\ref{asympt_region}), we infer that 
$z\phi_z \sim -\frac{t+r}{2}\phi_t(1+\frac{\phi_t}{2})$. Using this asymptotic 
relation in (\ref{as_phi_t}), we finally obtain the following result. 

\vskip 10pt
\noindent
In the space-time region 
\beq\label{asympt_region2}
\ba{l}
z=\frac{t+r}{2}e^{i\alpha},~~~~~~~\alpha, r\in\RR,~~~~~~t>>1, \\
X:=r+(t+r)\phi_t+\frac{3t}{8}\phi^2_t= \\
\sqrt{x^2+y^2}-t+\sqrt{x^2+y^2}\phi_t+\frac{3t}{8}\phi^2_t=O(1), 
\ea
\eeq
the longtime $t>>1$ behaviour of the solutions of the 
2ddT equation $(exp\phi_t)_{t}=\phi_{xx}+\phi_{yy}$ is described by the following implicit (scalar) equation:
\beq\label{as_sol}
\ba{l}
\phi_t=\frac{1}{\sqrt{t}}F\left(\sqrt{x^2+y^2}-t+\sqrt{x^2+y^2}\phi_t+\frac{3t}{8}\phi^2_t,\arctan \frac{y}{x}\right)+
o\left(\frac{1}{\sqrt{t}}\right), 
\ea
\eeq 
where $F$ is given by  
\beq\label{def_F}
\ba{l}
F\left(X,\alpha\right)=-\frac{1}{2\pi i}\int\limits_{\RR}d\mu'R_2\Big( -\frac{{\mu'}^2}{2}+X+a_1(\mu';X,\alpha) ,\alpha +   
a_2(\mu';X,\alpha)\Big) 
\ea
\eeq
and $a_j(\mu;X,\alpha),~j=1,2$ are the solutions of the integral equations
\beq\label{as_int_a}
\ba{l}
a_j(\mu;X,\alpha) = \frac{1}{2\pi i}
\int\limits_{\RR}\frac{d\mu'}{\mu'-(\mu - i\epsilon)}
R_j\Big( -\frac{{\mu'}^2}{2}+X+a_1(\mu';X,\alpha) ,\alpha + \\ 
a_2(\mu';X,\alpha)\Big),~~j=1,2.
\ea
\eeq
Outside the asymptotic region (\ref{asympt_region2}) the solution decays faster. 

We first remark that, since $\phi_t=O(t^{-\frac{1}{2}})$, the condition $X=O(1)$ implies that 
$r=\sqrt{x^2+y^2}-t=O(\sqrt{t})$;  
it follows that, in the longtime regime $t>>1$, the solution of the Cauchy problem for the 2ddT equation is 
concentrated, with amplitude $O(t^{-\frac{1}{2}})$, in the asymptotic region $\sqrt{x^2+y^2}-t=O(\sqrt{t})$. 
We also remark that the asymptotic solution (\ref{as_sol})-(\ref{as_int_a}) is connected to the initial conditions 
of the Chauchy problem through the direct problem presented in the previous section.    

%%%%%%%%%%%%%%%%%%%%%%%%%%%%%%%%%%%%%%%%%%%%%%%%%%%%%
\section{A distinguished class of implicit solutions}
%%%%%%%%%%%%%%%%%%%%%%%%%%%%%%%%%%%%%%%%%%%%%%%%%%%%%

In this section, in analogy with the results of \cite{MS5},\cite{BDM}, we construct a class of explicit solutions 
of the vector nonlinear RH problem (\ref{RH1}) and, correspondingly, a class of implicit solutions 
of the 2ddT equation parametrized by an arbitrary real spectral function of one variable. 

Suppose that the two components of the RH spectral data $\vec R$ in (\ref{RH1}) are given by:
\beq
R_j(s_1,s_2)=(-1)^{j+1} i f\left(e^{s_1+s_2}\right),~~j=1,2,
\eeq
in terms of the single real spectral function $f$ of a single argument, depending on 
$s_1$ and $s_2$ only through their sum. 

Then the RH problem (\ref{RH1b}) becomes   
\beq\label{RH3}
\ba{l}
\pi^+_1=\pi^-_1+if\left(e^{\pi^-_1+\pi^-_2}\right),~~~|\lambda |=1, \\
\pi^+_2=\pi^-_2-if\left(e^{\pi^-_1+\pi^-_2}\right) 
\ea 
\eeq
and the following properties hold.

\vskip 10pt
\noindent
i) The reality and Hamiltonian constraints (\ref{reality_R}) and (\ref{Ham_R}) are satisfied.

\vskip 10pt
\noindent
ii) $\pi^+_1+\pi^+_2=\pi^-_1+\pi^-_2$. Consequently, using the analyticity properties of the 
eigenfunctions, it follows that the functions $\Delta^{+}$ and $\Delta^{-}$, defined by
\beq
\Delta^{\pm}:=\pi^{\pm}_1+\pi^{\pm}_2-(z\lambda+\bar z\lambda^{-1})v-i\ln\lambda,
\eeq
are analytic respectively inside and outside the unit circle of the $\lambda$-plane 
and satisfy the equation $\Delta^+=\Delta^-$; therefore they are equal to a constant in $\lambda$. 
Evaluating such a constant at $\lambda=0$ and at $\lambda\to\infty$, we obtain the following equalities
\beq
\Delta^+=\Delta^-=-t-\bar z \phi_{\bar z}-i\frac{\phi_t}{2}=-t-z\phi_z+i\frac{\phi_t}{2}.
\eeq 
This implies that \\
i) the solutions of the 2ddT equation generated by the above RH problem satisfy the linear ($2+1$)-dimensional PDE
\beq\label{lin_constr}
\phi_t=i(\bar z \phi_{\bar z}-z\phi_z)
\eeq
and, substituting in (\ref{dT3}) the expression of $\phi_t$ in terms of $\phi_z,\phi_{\bar z}$ given in 
(\ref{lin_constr}), one obtains the following nonlinear two dimensional constraint: 
\beq
i(\bar z\partial_{\bar z}-z\partial_z)\left(e^{i(\bar z \phi_{\bar z}-z\phi_z)}\right)=\phi_{z\bar z}
\eeq
on the solutions of 2ddT constructed by the above RH problem. \\
ii) $\pi^+_1+\pi^+_2=\pi^-_1+\pi^-_2$ is the following explicit and elementary function of $\lambda$: 
\beq\label{def_w}
\ba{l}
w(\lambda):=\pi^+_1+\pi^+_2=\pi^-_1+\pi^-_2=(z\lambda+\bar z\lambda^{-1})e^{-\frac{\phi_t}{2}}+i\ln\lambda 
-t-z\phi_z+i\frac{\phi_t}{2}.
\ea
\eeq
iii) Since, from (\ref{def_w}), $\pi^{-}_1+\pi^{-}_2=w(\lambda)$ is an explicit function of $\lambda$, 
the vector nonlinear RH problem (\ref{RH3}) decouples into two scalar, linear RH problems:
\beq\label{RH_linear}
\ba{l}
\pi^+_1=\pi^-_1+ if\left(e^{w(\lambda)}\right), \\
\pi^+_2=\pi^-_2-if\left(e^{w(\lambda)}\right),
\ea
\eeq
whose explicit solutions are given by
\beq
\ba{l}
\xi^{\pm}_j(\lambda)=(-1)^{j+1}
\frac{1}{2\pi i}\oint_{|\lambda |=1}\frac{d\lambda'}{\lambda'-(1\mp\epsilon)e^{arg\lambda}}f\left(e^{w(\lambda')}\right)
, \ \ \ j=1,2,
\ea
\eeq
where $\xi^{\pm}_j=\pi^{\pm}_j-\nu_j$, and the closure conditions (\ref{closure1}) read
\beq\label{closure4}
\phi_z e^{-\frac{\phi_t}{2}}=-\frac{1}{2\pi i}\oint_{|\lambda |=1}d\lambda f\left(e^{w(\lambda)}\right), \ \ \ \
\phi_t = \frac{1}{2\pi i}\oint_{|\lambda |=1}\frac{d\lambda}{\lambda}f\left(e^{w(\lambda)}\right).
\eeq
Although the RH problem (\ref{RH_linear}) is linear, since $w(\lambda)$ in (\ref{def_w}) depends on the unknowns 
$\phi_t,\phi_z$, the closure conditions (\ref{closure4}) are a nonlinear algebraic system of two equations 
for the two unknowns $\phi_t,\phi_z$, defining implicitly a class of solutions of the 2ddT equation   
parametrized by the arbitrary real spectral function $f(\cdot )$ of a single variable. 

\vskip 10pt
\noindent
{\bf Acknowledgements}. This research has been supported by the RFBR 
grants 07-01-00446, 06-01-90840, and 06-01-92053, by the bilateral agreement 
between the Consortium Einstein and the RFBR, and by the bilateral agreement between the 
University of Roma ``La Sapienza'' and the Landau Institute for Theoretical Physics of the 
Russian Academy of Sciences.

\end{document}